\documentclass[twocolumn]{aastex631}
\usepackage{graphicx}
\usepackage{color}
\usepackage[T1]{fontenc}
\usepackage{amssymb,amsmath}
\usepackage{ulem}
\usepackage{url}
\usepackage{bm}
\newcommand{\HII}{H\,{\sc ii}}
\newcommand{\bfp}{\mbox{\boldmath $p$}}
\newcommand{\be}{\begin{equation}}
\newcommand{\ee}{\end{equation}}
\newcommand{\ba}{\begin{eqnarray}}
\newcommand{\ea}{\end{eqnarray}}

\received{September 4, 2021}
\revised{October 7, 2021}
\accepted{October 26, 2021}
\submitjournal{ApJL}

\shorttitle{Effect of Magnetic Field Dissipation on Primordial Li Abundance}
\shortauthors{Lu and Kusakabe}

\begin{document}

\title{Effect of Magnetic Field Dissipation on Primordial Li Abundance}

\correspondingauthor{Motohiko Kusakabe}
\email{kusakabe@buaa.edu.cn}

\author[0000-0002-9739-9577]{Yini Lu}
\affil{School of Physics, and International Research Center for Big-Bang Cosmology and Element Genesis, Beihang University, \\
  37, Xueyuan Road, Haidian-qu, Beijing 100083, People's Republic of China}

\author[0000-0003-3083-6565]{Motohiko Kusakabe}
\affil{School of Physics, and International Research Center for Big-Bang Cosmology and Element Genesis, Beihang University, \\
  37, Xueyuan Road, Haidian-qu, Beijing 100083, People's Republic of China}



\begin{abstract}
  The dissipation effects of primordial magnetic fields on the primordial elemental abundances were investigated. When a magnetic field reconnects, its energy is converted to the kinetic energy of charged particles, as observed for solar energetic particles arriving on earth. This accelerates the cosmic background nuclei, and energetic nuclei induce nonthermal reactions. A constraint on the dissipation is derived from a theoretical calculation of the nonthermal reactions during Big Bang nucleosynthesis. We found that observations of the Li and D abundances can be explained if 0.01--0.1 \% of the cosmic energy density was utilized for nuclear acceleration after the electron--positron annihilation epoch.
  Reconnections of such amplitudes of magnetic fields generate outgoing jets, the bulk velocity of which evolves to values appropriate for cosmic ray (CR) nuclear energies of 0.1--1 MeV necessary for successful CR nucleosynthesis.
Therefore, acceleration of cosmic background nuclei during the dissipation of primordial magnetic fields is a possible generation mechanism of soft CRs that has been suggested as a solution to the cosmic Li problem. Among the solutions suggested without exotic physics, only the dissipating magnetic field model suggested here explains observations of both low Li and high D abundances. Our results demonstrate that signatures of strong magnetic fields in the early universe have been observed in primordial elemental abundances.
\end{abstract}




\section{Introduction}\label{sec1}
Astronomical objects ranging over various scales, from planets to cosmological structures, are associated with magnetic fields \citep{Durrer:2013pga}. The origin of these magnetic fields is one of the most important questions of our universe, and constraints on cosmological magnetic fields are placed from many observations \citep{BICEP2:2017lpa,Yamazaki:2018gmr,Minoda:2018gxj,Jedamzik:2020krr,Katz:2021iou}, and cosmological \citep{2006Sci...311..827I} and astrophysical \citep{2005ApJ...633..941H} origins of Galactic magnetic fields have been investigated. If a magnetic field is generated during a postulated inflationary expansion of the universe \citep{1992ApJ...391L...1R} before the Big Bang, and pertains to the observed magnetic fields in galaxies and galaxy clusters, the magnetic field energy is constrained to be much less than (of the order of $10^{-5}$ of) the cosmic microwave background (CMB) energy \citep{BICEP2:2017lpa,Yamazaki:2018gmr,Katz:2021iou}. However, cosmic magnetic fields are also generated during electroweak phase transition \citep{Vachaspati:2020blt} and neutrino decoupling \citep{Dolgov:2001nv} before Big Bang nucleosynthesis (BBN). Primordial elemental abundances are the best probes for the magnetic fields generated after the Big Bang. These magnetic fields have small coherence lengths within the horizons at those times that correspond to $\sim 10^{-4}$ pc (phase transition) and $\sim 10$ pc (neutrino decoupling) in the present universe, far below the galaxy size. Such small-scale magnetic fields dissipate during later cosmic evolution \citep{Durrer:2013pga}. Therefore, they would affect physics in the early universe but probably have decayed and escaped from current astronomical observations.

The BBN offers the deepest reliable probe of the early universe based on well-understood standard model physics \citep{Pitrou:2018cgg}. The standard BBN (SBBN) theory predicts that $\sim25$ \% of the baryonic mass of the universe consists of $^4$He, $\sim0.004$ \% consists of deuterium, and $\sim 3 \times 10^{-9}$ consists of $^7$Li. In SBBN, the primordial plasma is assumed to be an ideal gas, where nonthermal nuclear reactions contribute negligibly to abundance evolution \citep{Voronchev:2012zz}. Nonthermal cosmic ray (CR) nucleosynthesis during BBN has been studied since \citet{Reno:1987qw,Dimopoulos:1987fz}. The most interesting possibility is that, if a low-energy CR component exists in the early universe, the reaction $^7$Be($p$,$p\alpha$)$^3$He\footnote{The expression A(a,b)B is used for the reaction A+a$\rightarrow$b+B.} reduces $^7$Be abundance \citep{Kang:2011vz,Kang:2018dsd}. The primordial $^7$Li abundance is inferred from observations of metal-poor stars (MPSs \citep{Sbordone:2010zi,Spite:1982dd}, but it is a factor 3--4 smaller than the yield of the SBBN model \citep{Hayakawa:2021jxf}. Because the primordial Li abundance predominantly originates from $^7$Be produced in the nucleosynthesis epoch that eventually decays into $^7$Li via electron capture, destruction of $^7$Be by CRs can solve the Li problem. We note that, if hard CRs involving the creation of protons, neutrons, and their antiparticles are assumed \citep{Reno:1987qw,Dimopoulos:1987fz}, no solution is found as a result of D overproduction \citep{Kusakabe:2014ola}.

The cosmic expansion enhanced by the magnetic field \citep{Greenstein1969} reduces primordial Li abundance because of the stronger destruction of $^7$Be via $^7$Be($n$,$p$)$^7$Li by more abundant neutrons \citep{Kawasaki:2012va}. In addition, if the magnetic field and temperature fluctuate such that the total energy density is homogeneous, the effects on abundance depend on the fluctuation pattern \citep{Luo:2018nth}. However, Li reduction under a magnetic field is constrained by large effects on the D and $^4$He abundances. 
Observations of solar energetic particles \citep{Reames:2013hma} provide important evidence of the particle acceleration process during magnetic reconnection \citep{2010RvMP...82..603Y}. If strong primordial magnetic fields dissipate and their energy triggers the generation of energetic nuclei in the early universe, nonthermal reactions are induced and elemental abundances are altered. In this Letter, we investigate the effect of magnetic dissipation on primordial abundances and provide the first numerical result of nonthermal nuclear reactions that take into account CR production via magnetic reconnection as in solar flares and the Coulomb energy loss process during CR propagation.

We adopt the natural units of $\hbar=k=c=1$ for the reduced Planck constant $\hbar$, the Boltzmann constant $k$ and the light speed $c$.

\section{Model}

We derive the steady-state spectra of CRs by adopting injection spectra from the observation of solar energetic particles, that is, solar CRs consisting of protons, electrons, $^2$H and $^3$H, and helium (namely $^3$He and $^4$He). In solar flares, magnetic reconnection \citep{2010RvMP...82..603Y} can release some energy in the form of kinetic energies of charged particles. Charged particles accelerated by the magnetic energy experience energy loss through Coulomb scattering off the background electrons and positrons in the early universe. In our model, strong magnetic fields and their dissipation are assumed. The gain in energy of the accelerated charged particles is proportional to the nuclear charge \citep{Reames:2013hma}. Moreover, collisions of energetic particles with background particles cause destruction or production of elements and affect primordial nuclear abundances.

Various time scales relevant to particle acceleration via cosmological magnetic reconnection are estimated in the Appendix. Reconnections of the large-scale magnetic fields on scales of $L \gtrsim 10$ km generate jets of accelerated plasma around reconnection regions in the early universe. The bulk velocity of the accelerated plasma under the resistive regime of the Sweet--Parker model \citep{1958IAUS....6..123S,1957JGR....62..509P} evolves from $v_\mathrm{out} \sim 1$ (cf. Equation (\ref{eq_conservation}) with small $l_\mathrm{jet}$ values) initially and to $v_\mathrm{out} \ll 1$ (Equation (\ref{eq_final_v})) finally.
Such bulk flows effectively enhance the reactivities of only nuclei because of the far larger enhancements of kinetic energies compared with those of electrons and photons. Collisions of jets with surrounding static background plasma trigger nonthermal nuclear reactions. Therefore, nonthermal nucleosynthesis caused by magnetic field reconnection proceeds over macroscopic scales, which is analogous to local collisions of fast hypernova ejecta with circumstellar matter rather than reactions of individual CRs with the interstellar medium occurring universally in the Galaxy.

  A typical CR nuclear energy per nucleon inside the jets is given by
  \begin{eqnarray}
  E &=&m_\mathrm{u} \frac{v_\mathrm{out}^2} {2}
  =0.419~\mathrm{MeV}/A \left( \frac{v_\mathrm{out}}{0.03} \right)^2,
  \end{eqnarray}
where
$m_\mathrm{u}$ is the atomic mass unit and
$A$ is the nuclear mass number. 
Then, if dissipations occur for magnetic fields with energy densities of the order of $10^{-4}$--$10^{-3}$ of the total energy density, CR nuclei accelerated via the dissipation have kinetic energies sufficient for the soft-CR nucleosynthesis investigated in this study (see Appendix).

\subsection{$^7$Be($p,p\alpha$)$^3$He cross section}
The threshold energy is $E_\mathrm{th}=1.586$ MeV. We calculate the resonant cross section by taking into account the second excited state of $^8$B with excitation energy $E_X=2.32$ MeV ($J^\pi=3^+$), which dominates at low temperatures during BBN. The decay width for the exit channel is calculated by assuming that a proton is emitted and escapes from the $^8$B compound nucleus with a relative energy of $< E-E_\mathrm{th}$ and that a remnant ($^4$He+$^3$He) spontaneously separates. We adopted the penetration factor corresponding to the relative energy $E-E_\mathrm{th}$ at which the proton penetrability is maximum. 
The Coulomb functions are calculated with a subroutine by Barnett\footnote{www.fresco.org.uk/programs/barnett/index.htm},  with an angular momentum $l=1$ between $p$ and $^7$Be assumed. For the exit channel, the reduced width was set to unity to evaluate the maximum possible effect of this reaction. The proton decay width of the entrance channel was fixed to the experimental total decay width \citep{Tilley:2004zz}. \citet{Kang:2011vz,Kang:2018dsd} adopted the $^2$H($p$, $pn$)$^1$H cross section as a function of $E-E_\mathrm{th}$ as a substitute for the $^7$Be($p$,$p\alpha$)$^3$He reaction as a trial. Because the Coulomb penetration factors differ between the two reactions, the cross section we derived improves upon the previous value.

\subsection{Experimental cross-sectional data}
Table \ref{tab1} lists the 31 nonthermal reactions included in our computation. Cross-sectional data are mainly adopted from Experimental Nuclear Reaction Data (EXFOR) \citep{ZERKIN201831}, and threshold energies for respective reactions are based on the Triangle Universities Nuclear Laboratory Nuclear Data Evaluation Project \footnote{https://nucldata.tunl.duke.edu/index.shtml}.

\begin{table}
  \caption{\label{tab1} Nonthermal reactions included in this work and references to their cross sections}
  \begin{ruledtabular}
    \begin{tabular}{l|l}
      Reactions & References \\
      \hline
        $^2$H($d$,$n$)$^3$He,~
        $^2$H($\alpha$,$\gamma$)$^6$Li    & \\
        $^3$He($\alpha$,$\gamma$)$^7$Be,~
        $^7$Li($\alpha$,$t$)$^8$Be        & \\
        $^7$Li($t$,$\alpha$)$^6$He,~
        $^3$H($d$,$n$)$^4$He              & \\
        $^2$H($d$,$p$)$^3$H,~
        $^2$H($p$,$\gamma$)$^3$He         & \\
        $^3$H($\alpha$,$\gamma$)$^7$Li,~
        $^3$H($\alpha$,$n$)$^6$Li         & \\
        $^3$He($t$,$d$)$^4$He,~
        $^3$He($t$,$np$)$^4$He            & \\
        $^6$Li($p$,$^3$He)$^4$He,~
        $^6$Li($t$,$p$)$^8$Li             & \citet{ZERKIN201831}
\\
        $^7$Li($\alpha$,$n$)$^{10}$B,~
        $^7$Li($d$,$p$)$^8$Li             & \\
        $^7$Li($d$,$t$)$^6$Li,~
        $^7$Li($^3$He,$t$)$^7$Be          & \\
        $^7$Li($p$,$\alpha$)$^4$He,~
        $^7$Li($p$,$n$)$^7$Be             & \\
        $^7$Li($t$,$n$)$^9$Be,~
        $^7$Be($d$,$n$)$^8$B              & \\
        $^7$Be($p$,$\gamma$)$^8$B,~
        $^2$H($p$,$np$)$^1$H              & \\  
        $^2$H($\alpha$,$\alpha n$)$^1$H   & \\
      \hline
      $^2$H($n$,$\gamma$)$^3$H            & \citet{ZERKIN201831}, \\
                                          & \citet{2006PhRvC..74b5804N} \\
      \hline
$^6$Li($\alpha$,$p$)$^9$Be,~
      $^6$He($\alpha$,$n$)$^9$Be      &\citet{2013ApJ...767....5K}    \\
      \hline
      $^6$Li($\alpha$,$d$)$^8$Be     &\citet{Fujiwara:1993nm}        \\
      \hline
      $^7$Be($p$,$p\alpha$)$^3$He     & this work                    \\
      \hline
$^7$Be($\alpha$,$p$)$^{10}$B   &\citet{Yamaguchi:2012sz}              \\
    \end{tabular}
  \end{ruledtabular}
\end{table}

\subsection{Source spectra case A (exponential cutoff)}
The following CR source spectrum was adopted from the observed spectra of solar energetic particles accelerated during solar magnetic field dissipation \citep{Reames:2013hma}:
\begin{equation}
Q_i^\mathrm{A}(E ;T) = Q_0^\mathrm{A}(T) Y_i E_\mathrm{MeV}^{-\gamma}
\exp \left(-E /E_{0i} \right),
\label{eq_s1}
\end{equation}
where $E$ is the nuclear kinetic energy per nucleon, $T$ is the temperature, $Y_i=X_i/A_i$ is the nuclear mole fraction, $E_\mathrm{0i}=E_0 (Z_i/A_i)$ is the cutoff scale, $X_i$ is the mass fraction, and $Z_i$ and $A_i$ are the charge and mass numbers of nuclide $i$, respectively. The subscript MeV indicates a quantity in units of MeV/$A$, and $Q_i^\mathrm{A}$ and $Q_0^\mathrm{A}$ have dimensions of cm$^{-3}$ s$^{-1}$ (MeV/$A$)$^{-1}$. The amplitude of the source spectrum is related to the total energy injection rate $\varepsilon^\mathrm{tot}$ as follows:
\begin{eqnarray}
\varepsilon^\mathrm{tot}(T) &=& \int_{E_\mathrm{min}}^\infty \sum_i A_i Q_i^\mathrm{A}(E;T) E dE \label{eq_total1} \\
&\approx&
(\mathrm{MeV})
\frac{Q_0^\mathrm{A}(T)}{(\mathrm{MeV}/A)^{-1}}
C(\gamma, E_0, E_\mathrm{min}(T); Y_\mathrm{p})~~~
\label{eq_cr1}
\end{eqnarray}
\begin{eqnarray}
C(\gamma, E_0, E_\mathrm{min}(T); Y_\mathrm{p}) \hspace{-5eM} && \nonumber \\
&\approx&
X_\mathrm{p} E_{0,\mathrm{MeV}}^{2 -\gamma} \Gamma \left(2 -\gamma, \frac{E_\mathrm{min}}{E_{0}} \right) \nonumber \\
 && + Y_\mathrm{p} \left( \frac{E_{0,\mathrm{MeV}}}{2} \right)^{2 -\gamma}
  \Gamma \left(2 -\gamma, \frac{2E_\mathrm{min}}{E_{0}} \right),~~~
  \label{eq_norm1}
\end{eqnarray}
where
$\Gamma(b,x) =\int_x^\infty t^{b-1} e^{-t} dt$ is the upper incomplete gamma function and 
$X_\mathrm{p}$ and $Y_\mathrm{p}$ are the mass fractions of $^1$H and $^4$He, respectively.
$C$ is a normalization constant, and we neglected contributions from CR nuclei, except for $^1$H and $^4$He because of their predominance in background abundance.
Typically, background charged particles have a kinetic energy of $~\sim T$. Therefore, the lower bound of the CR energy is taken as $E_\mathrm{min} =T$.

We adopt the parameterization given by
\begin{eqnarray}
\varepsilon^\mathrm{tot}(T) =\frac{f_\mathrm{dis} \rho_\mathrm{rad}(T)}{\Delta t},
\label{eq_budget1}
\end{eqnarray}
where the dimensionless parameter $f_\mathrm{dis}$ is the ratio of the total CR energy to the background radiation energy,
$\rho_\mathrm{rad}(T)$ is the background radiation energy density, and
$\Delta t$ is the duration of CR generation. In this study, we assume that CR generation operates from $T_9 \equiv T /(10^9~\mathrm{K})$=0.2 to 0.1.
In the standard cosmology \citep{Kolb:1990vq},  $T_9=0.2$ and 0.1 correspond to $t=4.4436 \times 10^3$ and $1.7774 \times 10^4$ s, respectively, and the duration is $\Delta t=1.33 \times 10^4$ s.
By using Equations (\ref{eq_cr1})--(\ref{eq_budget1}),
the amplitude of the CR source spectra is related by
\begin{eqnarray}
Q_0^\mathrm{A}(T) &=& 4.32 \times 10^{22}~\mathrm{cm}^{-3}~\mathrm{s}^{-1}~(\mathrm{MeV}/A)^{-1}
f_\mathrm{dis} \nonumber \\
&& \times C(\gamma, E_0, T; Y_\mathrm{p})^{-1}
\left( \frac{\rho_\mathrm{rad}(T)}{\mathrm{g~cm}^{-3}} \right)
\left( \frac{\Delta t}{1.3 \times 10^4~\mathrm{s}} \right)^{-1}. \nonumber \\
\end{eqnarray}

\subsection{Case B (sharp cutoff)}
As another case, we adopt a CR source spectrum given by
\begin{eqnarray}
Q^\mathrm{B}_i(E ;T) = Q_0^\mathrm{B}(T) Y_i \frac{E_\mathrm{MeV}^{-\gamma}}
{\exp \left[\left( E -E_{0i} \right) /a_\mathrm{dif} \right] +1},
\label{eq_s2}
\end{eqnarray}
where
$a_\mathrm{dif}$ is the diffuseness of the spectral cutoff.
This spectrum can accommodate a sharper cutoff than that in case A. The CR source amplitude was normalized using Equation (\ref{eq_budget1}), which is similar to case A.

\subsection{Steady-state CR spectra}
Scattering by abundant electrons and positrons in the early universe quickly thermalizes low-energy CR nuclei accelerated by magnetic fields. For slow CR nuclei, the Coulomb energy loss \citep{Reno:1987qw} is much faster than destruction by nuclear reactions. Therefore, the Coulomb loss process determines the shape of the spectra from the generated moment. The steady-state spectra are given by
\begin{equation}\label{10}
  n_{2i}^X(E) =\frac{1}{(dE/dt)^\mathrm{Coul}_i} \int_{E/A_i}^\infty Q_i^{X}(E'/A_i) d(E'/A_i),
\end{equation}
where $X$ = A or B depending on the source spectra $Q_i^X$,
$E$ is the kinetic energy of nuclei $i$, and
$(dE/dt)^\mathrm{Coul}_i$ is the Coulomb loss rate.
Note that this agrees with the limit of Coulomb loss dominance for the steady-state spectra of Galactic CRs \citep{2006A&A...448..665P}. The loss rate of nuclei in nonrelativistic $e^\pm$ background (i.e., $T \lesssim m_e$ where $m_e$ is the electron mass) is adopted from \citet{Reno:1987qw,Kawasaki:2004qu}.

\subsection{Nonthermal reaction rates}
Nonthermal reactions between a background nucleus (particle 1) and a CR nucleus (particle 2) were considered. The nonthermal reaction rates are given \citep{Kang:2011vz,Kang:2018dsd} by
\begin{eqnarray}\label{13}
  \langle \sigma v \rangle_{ij}^X(T; \bm{Z}^X)
  &=& \int_{-1}^1 d\mu \int_0^\infty dE_1 f_1(E_1;T) \nonumber \\
  && \times \int dE_2 f_{2j}^X(E_2;T; \bm{Z}^X) \sigma(v) v(E_1, E_2, \mu), \nonumber \\
\end{eqnarray}
where $\langle \sigma v \rangle^X$ is the average of the product of cross section $\sigma$ and relative velocity $v$ for $X$=A and B,
$f_1(E_1;T)$ is a Maxwell--Boltzmann distribution function for the energy of background nuclei, $E_1$,
$f_{2j}^X(E_2) =n_{2j}^X(E_2)/n_{2j,\mathrm{tot}}^X$
is the normalized nonthermal distribution function of the energy $E_2$ of CR nuclei $j$ with $n_{2j,\mathrm{tot}}^X =\int n_{2j}^X(E_2) dE_2$ being the total CR number density,
$\bm{Z}^X$ are parameters of the CR source spectra ($\bm{Z}^\mathrm{A} =(\gamma, E_\mathrm{0})$ and $\bm{Z}^\mathrm{B} =(\gamma, E_\mathrm{0}, a_\mathrm{dif})$), and 
$\mu =\cos \theta$ is the cosine of the incidence angle $\theta$.
We define the $G$-function as
\begin{equation}\label{15}
 \begin{split}
G(E_2; T) =
\int_0^\infty dE_1 f_1(E_1;T)
\int_{-1}^1 d\mu
\left[ \sigma v \right] (E_1, E_2, \mu).
 \end{split}
\end{equation}
This is the reaction rate of a CR with energy $E_2$ at a background temperature $T$, which is independent of the CR source spectrum. By utilizing this generic quantity, the integrated reaction rate can be written as
\begin{eqnarray}\label{13}
  \langle \sigma v \rangle^X(T; \bm{Z}^X)
  &=& \int dE_2 f_{2j}^X(E_2;T; \bm{Z}^X)
  G(E_2; T).~~~
\end{eqnarray}

\subsection{BBN calculation}
We adopted the SBBN code NUC123 \citep{Kawano:1992ua,Smith:1992yy} and updated the reaction rates of nuclei with mass numbers of $\le  10$ using the JINA REACLIB database \citep{2010ApJS..189..240C} (updated on May 14, 2021) and reaction rates of $^2$H($p$,$\gamma$)$^3$He, $^2$H($d$,$n$)$^3$He, and $^2$H($d$,$p$)$^3$H \citep{Coc:2015bhi}. In addition, $^6$He was included as a new nuclear species with nuclear data from \citet{Wang:2021xhn}, and a new reaction type, that is, A+B$\rightarrow$C+D+E, was encoded. The baryon-to-photon ratio was set to $\eta =6.133 \times 10^{-10}$ from the Planck CMB power spectra, CMB lensing, and baryon acoustic oscillation for the base-$\Lambda$CDM model, $\Omega_\mathrm{b} h^2 = 0.0224 \pm 0.0001$ \citep{Aghanim:2018eyx}. The neutron lifetime is the central value of $\tau =879.4 \pm 0.6$ s \citep{Zyla:2020zbs}.

\section{Results}
\begin{figure}[h]
  \centering
  \includegraphics[width=3.5in]{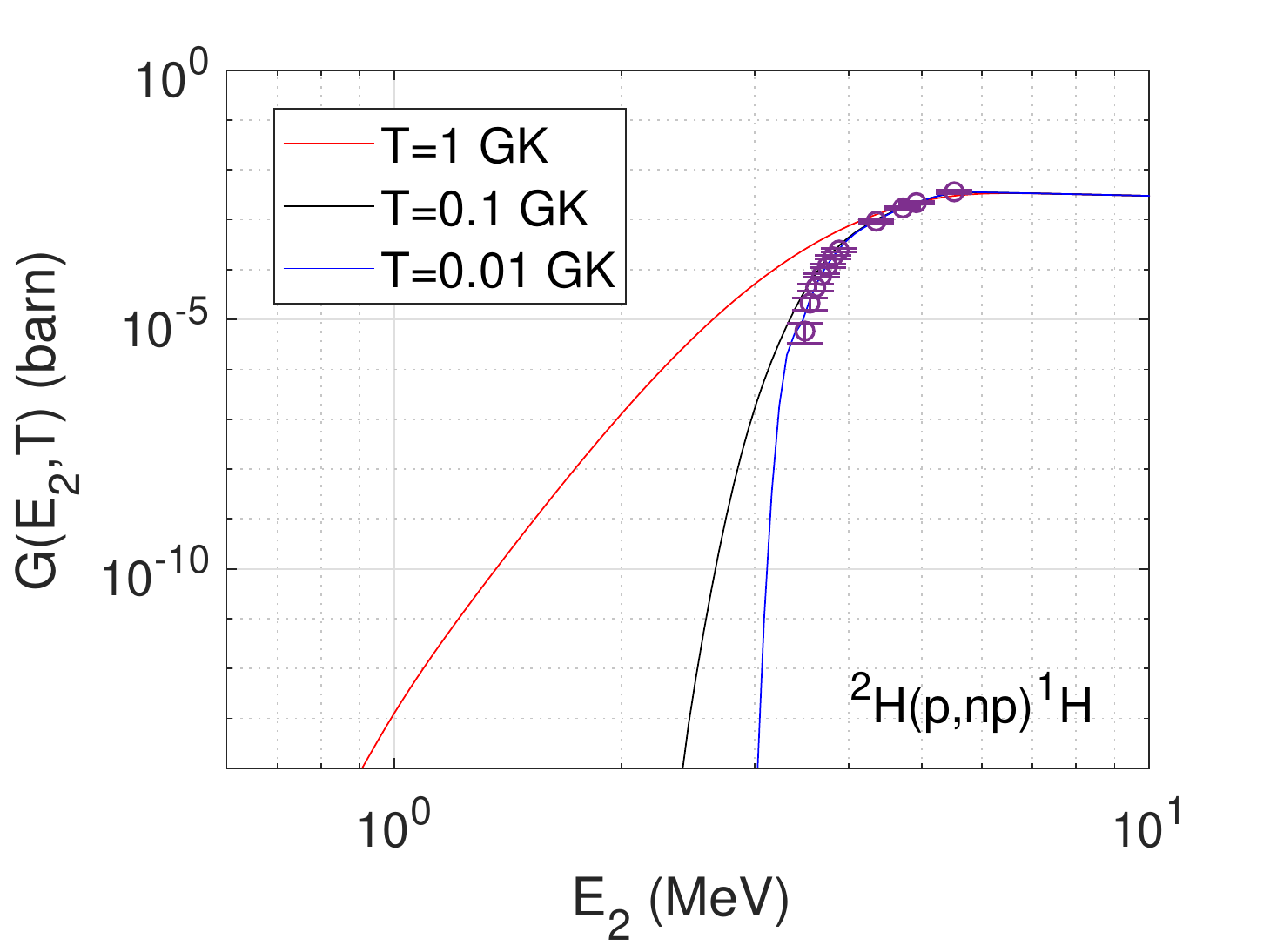}\\
  \includegraphics[width=3.5in]{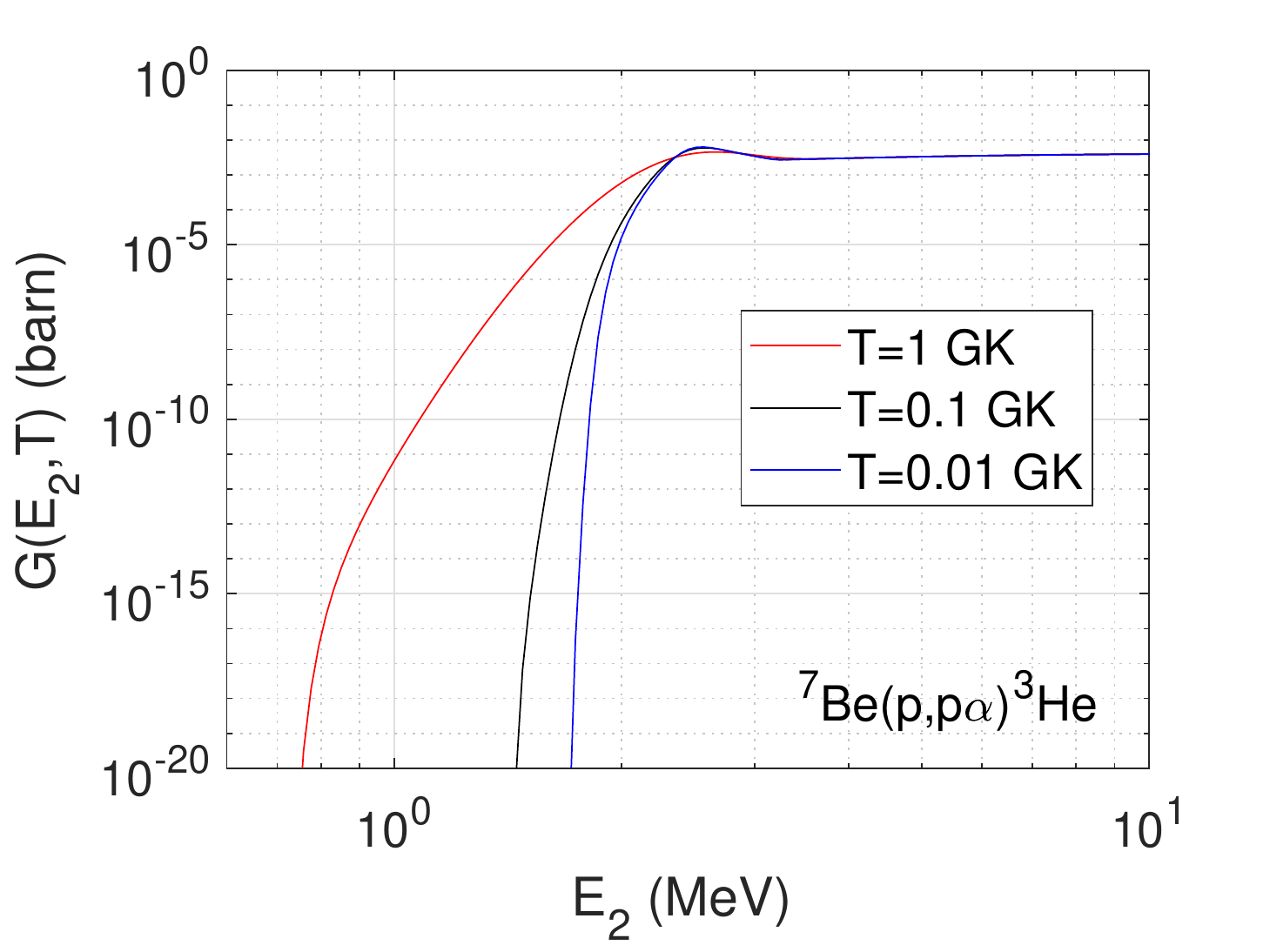}\\
  \caption{Nonthermal rates for the reactions $^2$H$(p,np)^1$H (top) and $^7$Be$(p,p\alpha)^3$He (bottom) as a function of the CR proton energy $E_2$ for cosmic temperature $T=1$, 0.1, and 0.01 GK, respectively. At high temperatures, less-energetic protons below threshold energies can react because of energetic background nuclei in the tail of the Maxwell--Boltzmann distribution. Also shown are data points of $\sigma v$ values \citep{Gibbons:1959zz} that correspond to $T=0$.}\label{1}
\end{figure}

Figure \ref{1} shows the nonthermal reaction rates as a function of the CR energy $E_2$ at fixed temperatures, that is, $G(E_2;T)$ (Eq. \ref{15}). The rates are the average reaction rates weighted by the background nuclear distribution. A finite temperature background effect was found as the difference between various temperatures. The $^2$H$(p,np)^1$H and $^7$Be$(p,p\alpha)^3$He reactions have positive threshold energies, and their cross sections are zero below the thresholds when the energy of the background target nucleus is neglected. However, at high temperatures, there are abundant energetic nuclei in the background, which help the reactions of low-energy CRs. Therefore, the nonthermal reaction rates at low-$E_2$ levels are significantly enhanced at high temperatures. We note that this finite temperature effect on CR reaction rates has not been considered in the Galactic CR nucleosynthesis calculations \citep{2006A&A...448..665P}. Although the effect is small for low background temperatures in Galactic interstellar matter and high typical CR energies of the order of 0.1--1 GeV, it affects nonthermal nucleosynthesis by soft CR nuclei in the early hot universe, as investigated in this study.

\begin{figure}[h]
  \centering
  \includegraphics[width=3.5in]{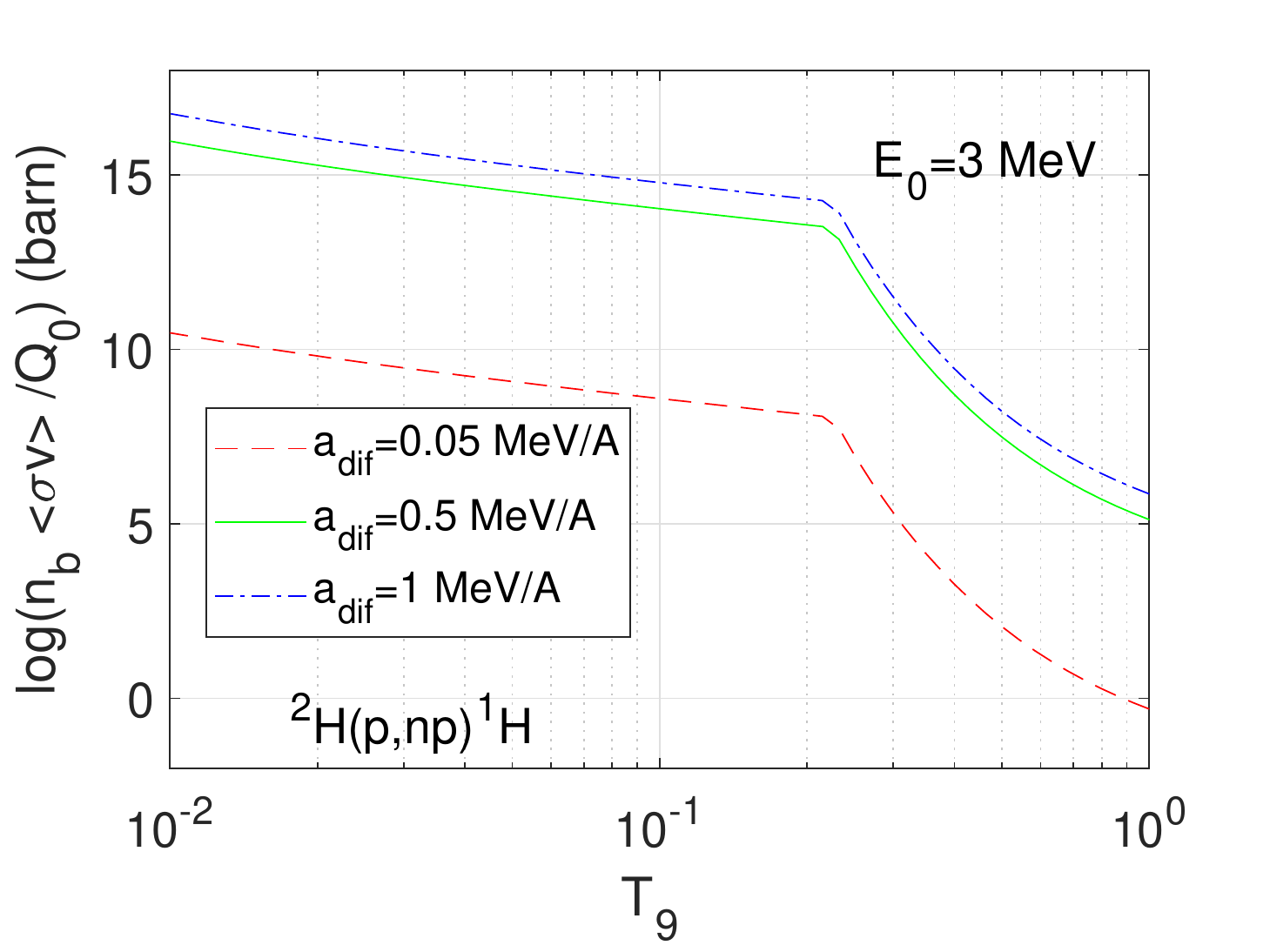}\\
  \includegraphics[width=3.5in]{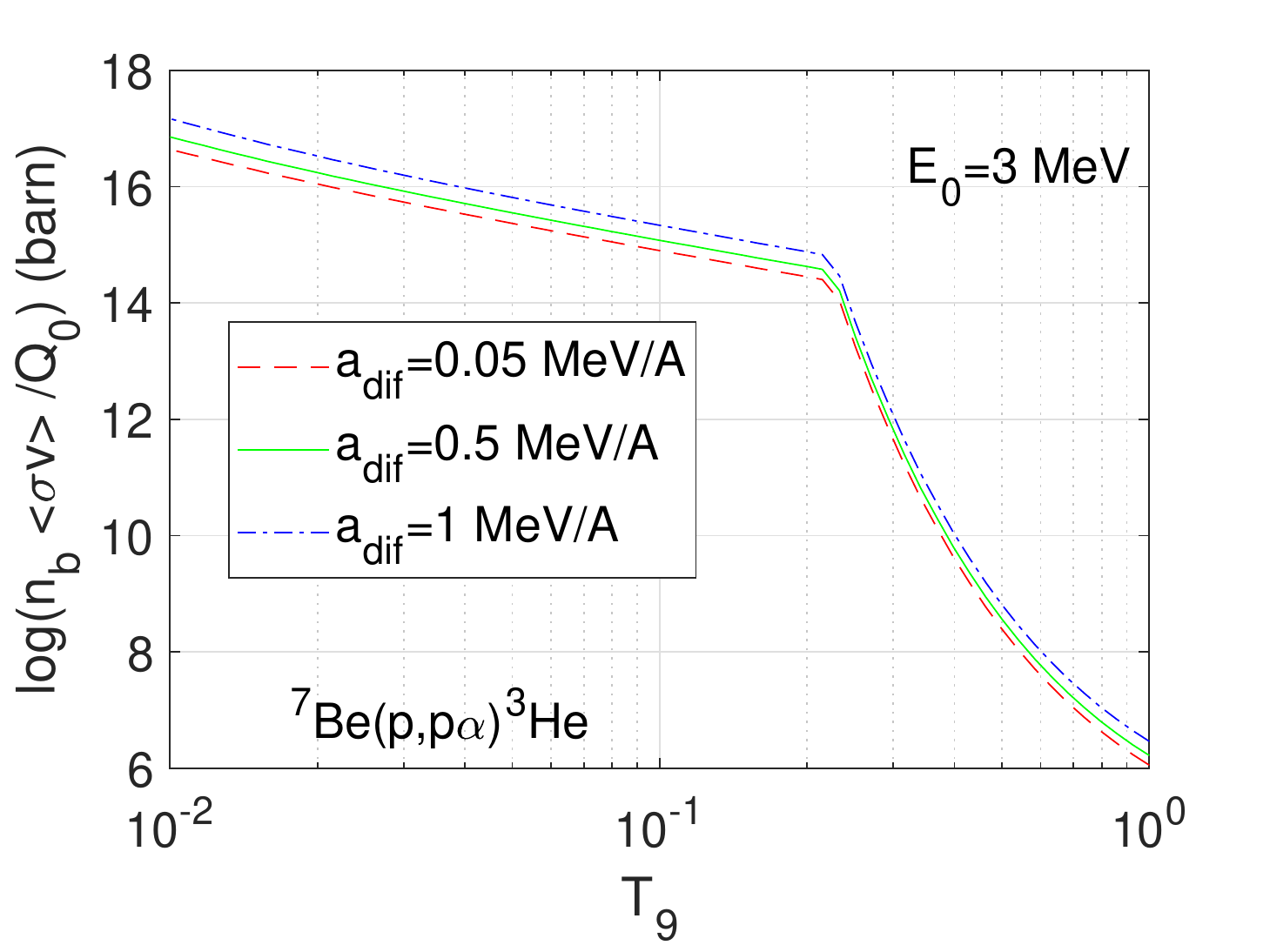}\\
  \caption{Integrated rates of the reactions $^2$H$(p,np)$$^1$H (top) and $^7$Be$(p,p\alpha)$$^3$He (bottom) as a function of temperature $T_9=T/(10^9~\mathrm{K})$. For the rates, $n_b$ is the baryon number density. It is assumed that the CR source spectrum of protons has a power-law index $\gamma=0$ and a sharp cutoff at $E_0=3$ MeV with diffuseness values of  $a_\mathrm{dif}=0.05$, 0.5, and 1 MeV, respectively (case B).
  }\label{2}
\end{figure}

Figure \ref{2} shows the reaction rates of the CR protons for $^2$H$(p,np)$$^1$H and $^7$Be$(p,p\alpha)$$^3$He reactions in case B. The adopted cutoff scale $E_0=3$ MeV for proton energy is higher than the threshold energy of $^2$H destruction (3.337 MeV) and lower than that of $^7$Be (1.813 MeV). Therefore, the reaction rate for the $^2$H($p,np$)$^1$H reaction is more sensitive to the sharpness of the cutoff. The nonthermal reaction rates monotonically decrease with increasing temperature for $T < m_e /26$ because the Coulomb energy loss rate increases. However, above the critical temperature $T = m_e /26$, the rates decrease suddenly with increasing temperature because electrons and positrons gradually become relativistic, and their number densities increase exponentially \citep[Appendix in][]{Reno:1987qw}. This figure clarifies that nonthermal reactions triggered by magnetic field dissipation are effective after the critical temperature corresponding to the completion of electron--positron annihilation. In the temperature range of $m_e /26 \leq T \lesssim 1$ GK, the electron-to-baryon ratio decreases by nine orders of magnitude. Before the end of annihilation, nonthermal nuclei quickly lose energy via scattering off of abundant electrons and positrons. The resulting low reaction rates lead to minor effects on elemental abundances. However, after the end of annihilation, a small number of electrons result in large amplitudes of the steady-state spectra of CR nuclei. Therefore, nonthermal reactions effectively affect elemental abundances. We note that the magnetic field energy per baryon decreases with decreasing temperature because of the dilution of the magnetic field. This reduces the effects of nonthermal reactions at low temperatures. In the following, we concentrate on the epoch of effective nonthermal reactions. For example, we adopt a case of CR generation in the range of $T_9=0.2$--0.1.

The results of nonthermal nucleosynthesis calculations were compared to $2\sigma$ observational constraints adopted as follows: $^4$He abundance $Y_p=0.2453 \pm 0.0034$ in \HII~regions in metal-poor galaxies \citep{2021JCAP...03..027A}, 
D abundance D/H= $(2.545 \pm 0.025)\times10^{-5} $ in Lyman-$\alpha$ absorption systems of quasar emissions \citep{2018MNRAS.477.5536Z},
$^3$He abundances $^3$He/H=$(1.9\pm 0.6)\times 10^{-5}$ in Galactic \HII~regions~\citep{Bania:2002yj} (where only the upper limit is taken), 
$^7$Li abundance log($^7$Li/H) = $-12+(2.199\pm0.086)$ in Galactic MPSs \citep{Sbordone:2010zi}, 
 and $^6$Li abundance $^6$Li$/$H$ = (0.85 \pm 4.33) \times 10^{-12}$ (where only the upper limit is taken) in Galactic MPS G64-12 \citep{Lind:2013iza}. For D abundance, our SBBN result is below the $2 \sigma$ limit, and the investigated parameter spaces do not have a 2$\sigma$ allowed region. Therefore, regions with 5\% and 10\% destruction of D are shown instead.

\begin{figure}[h]
  \centering
  \includegraphics[width=3.5in]{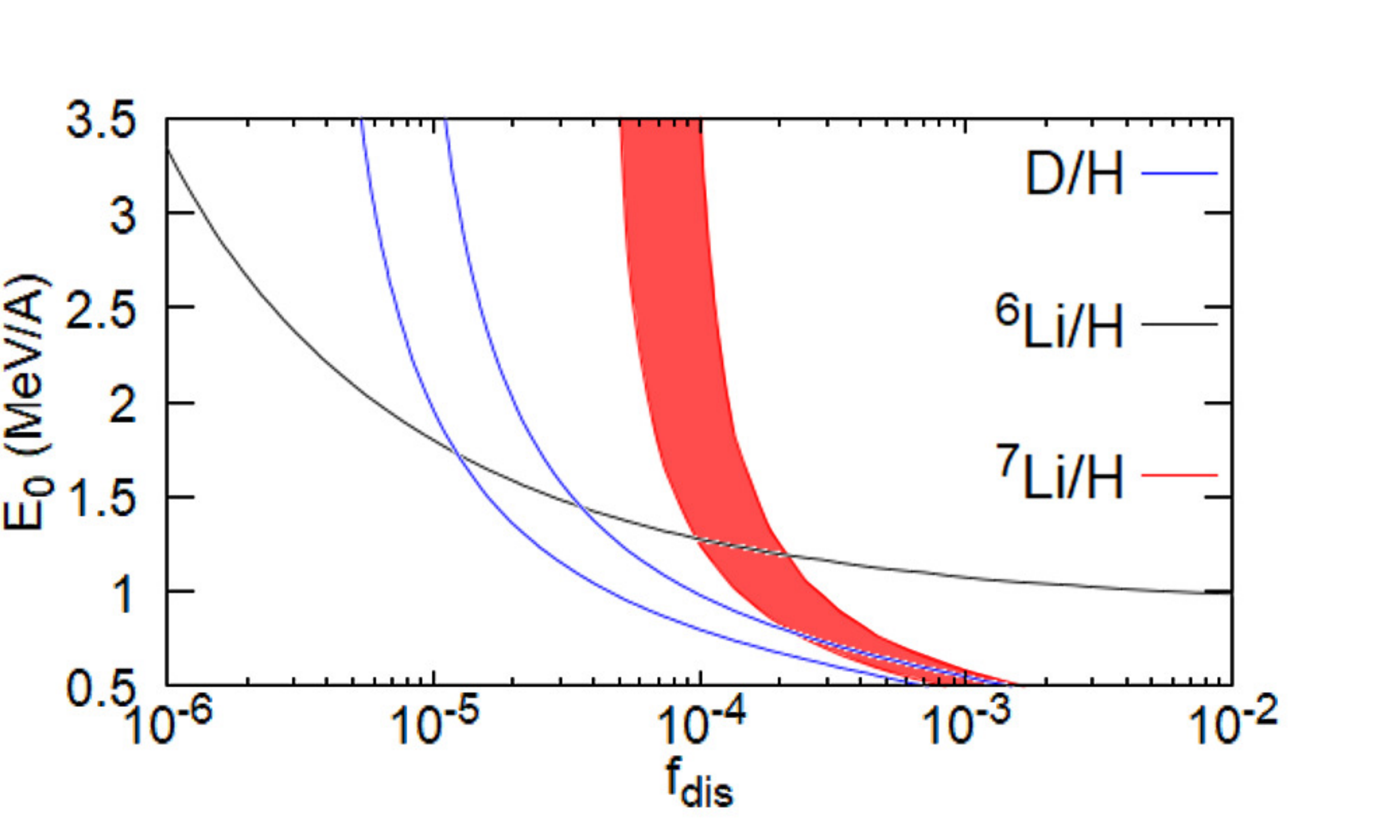}\\
  \includegraphics[width=3.5in]{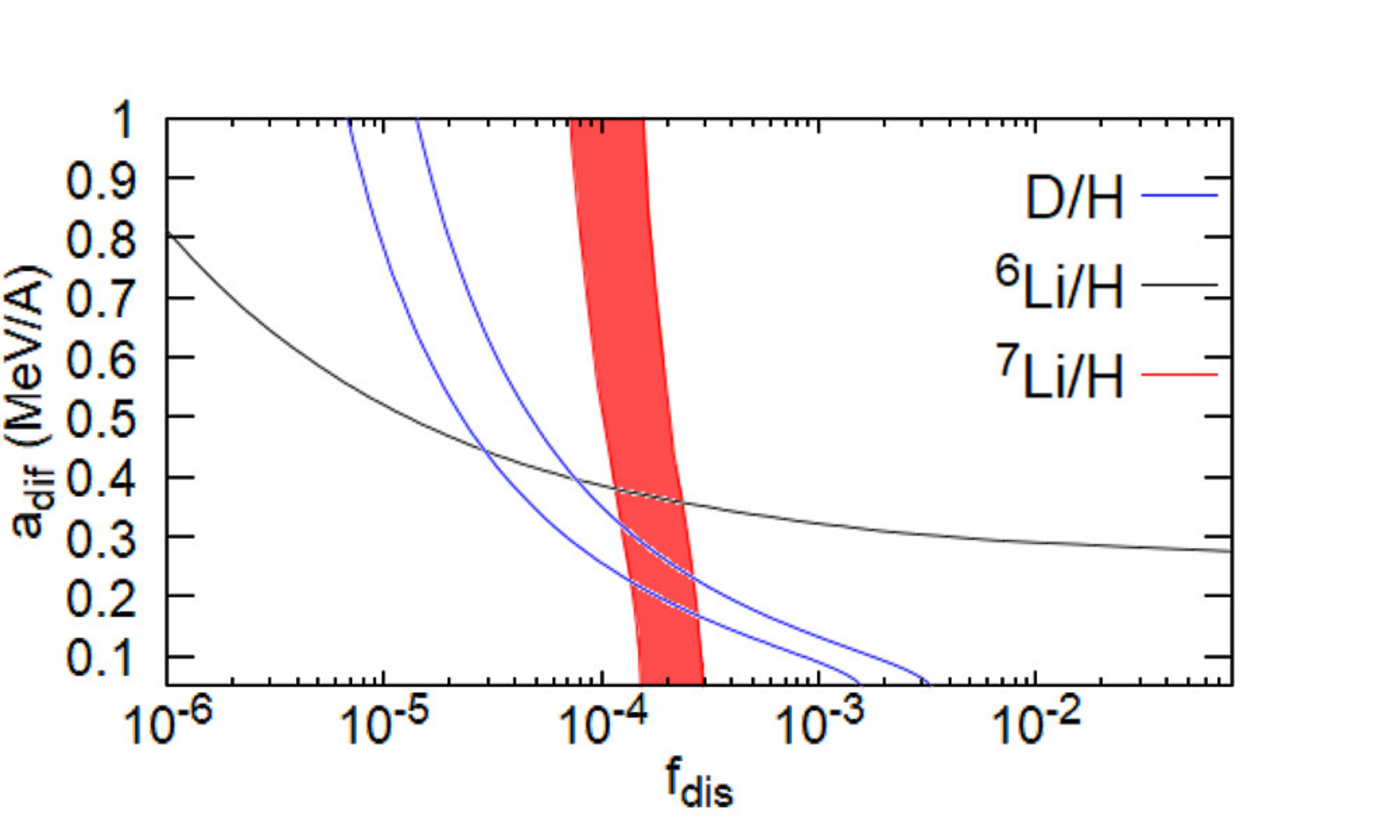}\\
  \caption{Contours of calculated primordial abundances in the parameter planes of ($f_\mathrm{dis}$, $E_0$) (case A, top) and ($f_\mathrm{dis}$, $a_\mathrm{dif}$) (case B, bottom). In the colored regions, the $^7$Li abundance agrees with observations of MPSs \citep{Sbordone:2010zi}. The lower and upper curves for D/H correspond to 95\% and 90\% of standard model value, respectively. Above the upper curve, significant destruction of D occurs. In the regions above the line for $^6$Li, the $^6$Li abundance is more than the upper limit from observations of MPSs \citep{Lind:2013iza}. A solution to the Li problem is located at the lower region inside the colored bands below the curves of D/H.}
  \label{3}
\end{figure}

Constraints on the magnetic dissipation are shown for cases A and B in Fig. \ref{3}. $^7$Be destruction in the early universe leads to lower primordial Li abundance after unstable $^7$Be decays into $^7$Li. The $^7$Li abundance after this decay is consistent with the observations in the colored bands. Areas to the right of the colored regions are excluded by a $^7$Li abundance that is too low. In contrast, areas to the left of the colored regions are still possibly allowed, although the $^7$Li abundance is higher than the observed level. After the BBN epoch, the $^7$Li abundance may be altered by the development of inhomogeneity in $^7$Li$^+$ ionic abundance during structure formation \citep{Kusakabe:2014dta} or Li depletion during the pre-main sequence \citep{2015MNRAS.452.3256F} and stellar evolution \citep{Korn:2006tv} of the observed MPSs. Therefore, it is also possible that nonthermal nuclear reactions are partially responsible for the Li problem. This case is located on the left side of the colored bands.

The $^6$Li abundance is higher than the observational upper limit above the loosely inclined lines. Although $^6$Li is fragile against nuclear burning via $^6$Li($p$,$\alpha$)$^3$He in stars, primordial $^6$Li abundance elevated by CRs above the upper limit may be observable in the near future by spectroscopic observations of MPSs. Above the $^6$Li lines, $^6$Li abundance rapidly increases with the cutoff scale (case A) and diffuseness of the cutoff (case B) of the CR source spectra. Therefore, areas far from these lines are excluded. The D abundance is significantly lower than that in the standard model above the two lines for D/H. The lower and upper lines correspond to 5\% and 10\% reductions, respectively, of the D/H value in the SBBN model. Regions above these lines are excluded from the overdestruction of D. We note that the existence of a magnetic field during BBN affects abundance evolution mainly through an increased cosmic expansion rate \citep{Greenstein1969}. For example, the primordial D/H abundance is most sensitively increased by 13\% if the magnetic field energy amounts to 13\% of total radiation energy during thermal nucleosynthesis operating at $T \gtrsim 1$ GK \citep{Kawasaki:2012va}. This effect works in the opposite direction of nonthermal nucleosynthesis and can be responsible for the high observed value of primordial D/H. Currently, a discrepancy in D abundance between the observations and the SBBN model is suspected \citep{2018MNRAS.477.5536Z}. This indicates the possibility that observations of the abundances of both $^7$Li and D have already revealed magnetic field effects in the early universe. In both cases A and B, below the bound from D/H, CR nucleosynthesis predicts nuclear abundances that fall into the observationally allowed ranges. These regions allow us to solve the problem of Li. It is found that $\sim$0.01\%--0.1 \% of cosmic energy density is needed for acceleration of background nuclei and that CR source spectra must have a sharp cutoff below the threshold energy of D spallation.

\begin{figure}[h]
  \centering
  \includegraphics[width=3.5in]{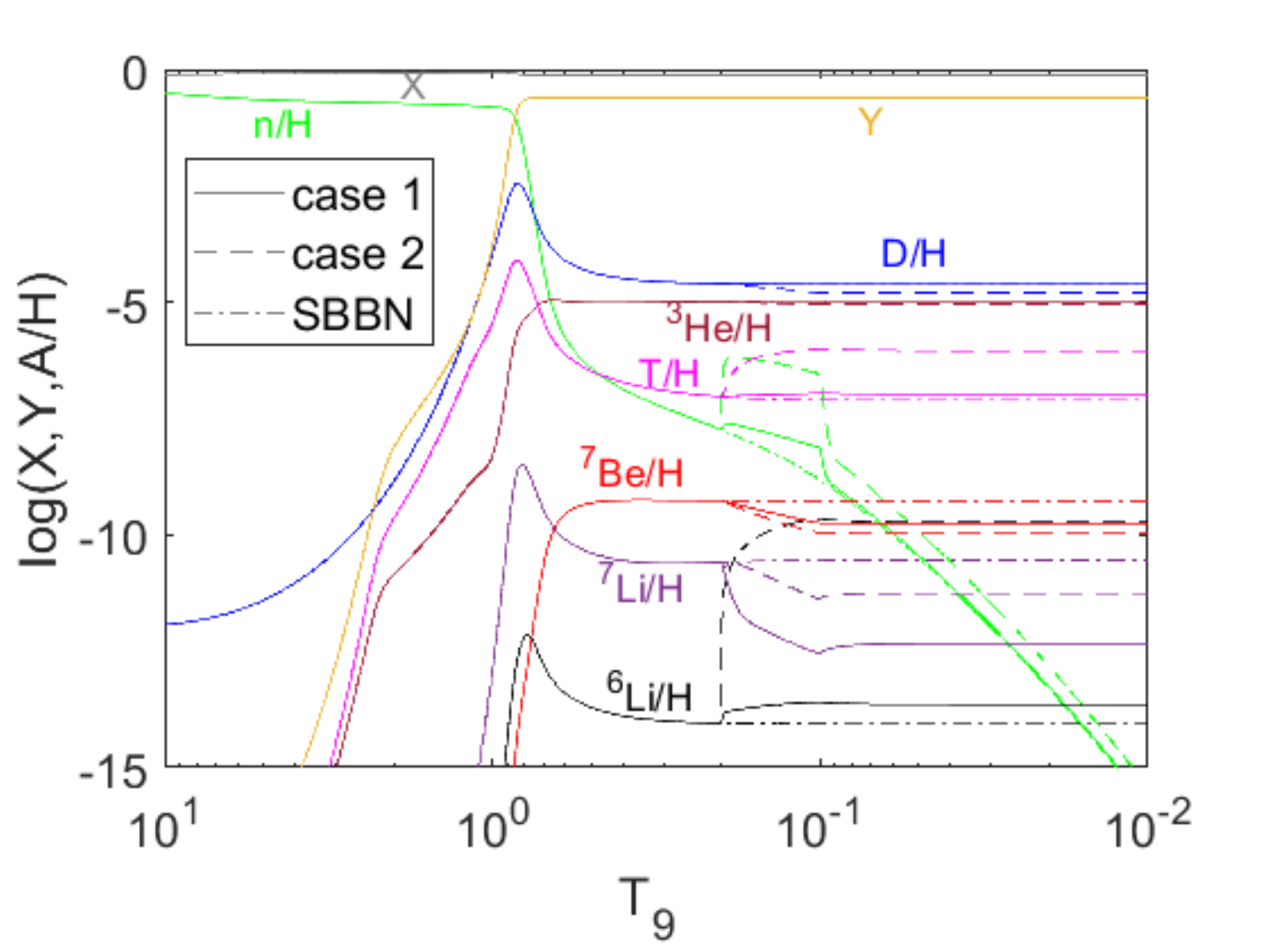}\\
  \caption{Nuclear abundances as a function of cosmic temperature $T_9=T/(10^9~\mathrm{K})$. Solid and dashed lines correspond to ($f_\mathrm{dis}$, $a_\mathrm{dif}$) = ($2\times 10^{-4}$, $0.1$) and ($2\times 10^{-4}$, $0.5$), respectively, in case B for magnetic field dissipation, while dash-dotted lines correspond to the standard model without a magnetic field. Cosmic ray generation at $T_9=0.2$--0.1 results in nonthermal nuclear reactions.
  }
  \label{4}
\end{figure}

The effects of magnetic field dissipation in case B on the evolution of the elemental abundances are shown in Fig. \ref{4}. In the SBBN model \citep{Pitrou:2018cgg}, abundances freeze out at $T\lesssim 1$ GK. However, if nuclear accelerations are realized from magnetic field dissipation, the abundances of $^7$Be, $^6$Li, and D evolve at the dissipation time. It was confirmed that $^7$Be and D are disintegrated via $^7$Be$(p,p\alpha)$$^3$He and $^2$H$(p,np)^1$H, respectively, and $^6$Li is predominantly produced via $^3$H$(\alpha,n)^6$Li. If an appropriate amount of energy is used for CR acceleration, primordial Li abundance can decrease to the abundance level of MPSs. However, the CR source spectrum must have a sharp cutoff. Otherwise, energetic protons excessively destroy D, and energetic tritons produce $^6$Li.

\section{Conclusions}\label{sec4}
Nonthermal nuclear reactions induced by soft CRs originating from magnetic field dissipation in the early universe were investigated. We adopted two types of CR source spectra with reference to the observed spectra of solar energetic particles energized by solar magnetic fields. Our findings are as follows: 
The CR flux can be significantly high only after the completion of $e^\pm$ pair annihilation. Therefore, nonthermal nucleosynthesis affects primordial abundance after the annihilation epoch. 
The magnetic dissipation after the annihilation can explain the observations of the primordial Li and D abundances if the dissipated energy amounts to 0.01\%--0.1\% of the total cosmic energy. 
The CR source spectra should be very soft, so that D overdestruction is not triggered. 
When we assume that reconnection of large-scale magnetic fields with energy densities of 0.01\%--0.1\% of the total energy density generates jets of tightly coupled electron--nuclei plasma, the kinetic energies of the CR nuclei in the jets evolve and can temporarily match the order of 1 MeV/$A$ required to solve the Li problem. 


%



\software{NUC123 \citep{Kawano:1992ua}, \\
  Fresco (www.fresco.org.uk/programs/barnett/index.htm)}

\appendix
\section{Momentum transfer rates and magnetic reconnection rate}

We derive rough estimates of momentum transfer rates at $T_9 =T /(10^9~{\rm K}) =0.1$ with $T$ the temperature, after the electron annihilation in the Big Bang. The quantity $\sigma_\mathrm{mt}^{ab}$ defines the momentum transfer cross section (MTCS) of particle $a$ via the scattering off of particle $b$.

\subsection{Electrons}
The momentum exchange of electrons is dominated by Coulomb scatterings with $e^-$s in the early universe.
The MTCS is then estimated considering scatterings off of target $e^-$ particles at rest in frame of the cosmic fluid under the nonrelativistic approximation.
The maximum impact parameter $b_\mathrm{max}$ is set to the Debye length $\lambda_{\rm D}$ as
\begin{eqnarray}
  b_{\rm max} &=&\lambda_{\rm D} = 4.58 \times 10^{-3}~{\rm cm}
  ~({\rm at}~T_9 = 0.1) \\
  \lambda_{\rm D} &\approx &\sqrt{ \frac{T}{4 \pi e^2 \left( n_{e} +n_\mathrm{b} \right)}} ~({\rm for}~T \lesssim m_e/26) \\
    &=&\sqrt{ \frac{\pi}{8 \zeta(3) \alpha \eta \left( 2 -Y_\mathrm{p}/2 \right) T^2 }},
\end{eqnarray}
where
$e$ and $m_e$ are the electronic charge and mass, respectively,
$n_e$ and $n_\mathrm{b}$ are the electron and baryon number densities, respectively,
$X_\mathrm{p} =0.75$ and $Y_\mathrm{p} =0.25$ are primordial mass fractions of $^1$H and $^4$He, respectively,
$n_\gamma =[2 \zeta(3) /\pi^2] T^3$ is the photon number density
with $\zeta(3) =1.2021$,
$\eta =n_\mathrm{b}/n_\gamma=6 \times 10^{-10}$ is the baryon-to-photon ratio, and
$\alpha=e^2$ is the fine structure constant.

The scattering angle $\theta$ is related to the impact parameter $b$ by
\begin{equation}
  b =\frac{q_1 q_2 e^2}{m v^2} \cot
  \left( \frac{\theta}{2} \right),
  \label{eq6}
\end{equation}
where
$q_1$ and $q_2$ are charge numbers of reacting particles 1 and 2, respectively,
$m=m_e/2$ is the reduced mass, and
  $v$ is the $e^-$ velocity
Then, the maximum of $\mu \equiv \cos{\theta}$ is given by
\begin{eqnarray}
  \mu_{\rm max} &=& \cos \theta_{\rm min} \\
  &=&
  \cos \left[ 2 \cot^{-1} \left( b_{\rm max} \frac{m_e v^2}{2 \alpha} \right) \right] \\
  & \approx& 1 -3.04\times 10^{-20} v^{-4}~~(\mathrm{at}~T_9=0.1).
  %
\end{eqnarray}
With this maximum $\mu_{\rm max}$, the MTCS is given by
\begin{eqnarray}
  \sigma_{\rm mt}^{ee} &\approx &\int_{-1}^{\mu_{\rm max}} \frac{d \sigma}{d\mu} \frac{\Delta q_e}{q_e} d\mu
 \\
  &\approx&
  \frac{2^{1/2}\pi \alpha^2}{v^2 q_e^2}
  \left[ 2 (1-\mu_{\rm max})^{-1/2} \right] \\
  &\approx &
  2.71 \times 10^6 /(v q_e)^2,
  \label{eq10}
\end{eqnarray}
where
$q_e$ and $\Delta q_e =2 q_e \sin(\theta/2)$ are the initial momentum and the momentum transfer of $e^-$, respectively,
and
the differential Mott cross section is given by
\begin{eqnarray}
  \frac{d \sigma}{d\mu}(v,\mu) &=&
  \frac{2\pi \alpha^2}{v^2 {q_e}^2}
  \frac{1}{(1-\mu)^2}
  \left[ 1 -\frac{v^2}{2} ( 1-\mu) \right].
  \label{eq_sigma_tot4}
\end{eqnarray}
The momentum transfer rate of electrons via the Coulomb scattering is then given by
\begin{eqnarray}
  \Gamma_{\rm C}^{ee} &\approx& n_{e} \sigma_{\rm mt}^{ee} \\
  &\sim & \eta n_\gamma (1 -Y_\mathrm{p}/2)
  \left[ \frac{2.71 \times 10^6}{(m_e v^2)^2} \right] \\
  &= & 1.29 \times 10^{9}~{\rm s}^{-1} v^{-4}~({\rm at}~T_9 =0.1).~~~~~
\end{eqnarray}

\subsection{Nuclei}
Scatterings are dominated by the Coulomb scattering with $e^-$. 
The maximum $\mu_\mathrm{max}$ is derived from Equation (\ref{eq6}) to be
\begin{eqnarray}
  \mu_{\rm max}
  &=&
  \cos \left[ 2 \cot^{-1} \left( b_{\rm max} \frac{m_e v^2}{Z \alpha} \right) \right] \\
  &\approx& 1 -7.61\times 10^{-21} Z^2 v^{-4}~~(\mathrm{at}~T_9=0.1),
\end{eqnarray}
where
$Z$ is the atomic number of the nucleus.
The MTCS of the nucleus is then given by
\begin{eqnarray}
  \sigma_{\rm mt}^{Ae} &\approx &\int_{-1}^{\mu_{\rm max}} \frac{d \sigma}{d\mu} \frac{\Delta q_e}{q_e} \frac{q_e}{q_A} d\mu
\\
  &\approx &
  \frac{2^{3/2} \pi Z^2 \alpha^2}{v^2 q_e q_A}
  \left[ 2 (1-\mu_{\rm max})^{-1/2} \right] \\
  &\approx &
  1.08 \times 10^7 Z/(q_e q_A),
  \label{eq_sigma_tot6}
\end{eqnarray}
where
$q_A$ is the nuclear momentum.
Accordingly, the momentum transfer rate is given by
\begin{eqnarray}
  \Gamma_{\rm C}^{Ae} &\approx& n_{e} \sigma_{\rm mt}^{Ae} \\
  &\approx &
  \eta n_\gamma (1 -Y_\mathrm{p}/2)
  \left[ \frac{1.08 \times 10^7 Z }{(m_e v) \sqrt{2 m_A E_A}} \right] \\
  &= & \frac{1.69 \times 10^{9}~{\rm s}^{-1} Z}{A^{1/2} v}
  \left(\frac{E_A}{3T/2} \right)^{-1/2}
  ~({\rm at}~T_9 =0.1),~~~~~
\end{eqnarray}
where
$A$ is the nuclear mass number, and
  $E_A$ is the nuclear kinetic energy.

The MTCS of electrons via scattering off of nuclei is, on the other hand, given by
\begin{eqnarray}
  \sigma_{\rm mt}^{eA} &\approx &\int_{-1}^{\mu_{\rm max}} \frac{d \sigma}{d\mu} \frac{\Delta q_e}{q_e} d\mu
\\
  &\approx &
  \frac{2^{3/2} \pi Z^2 \alpha^2}{v^2 q_e^2}
  \left[ 2 (1-\mu_{\rm max})^{-1/2} \right]
  \label{sigma_mt_ep0} \\
  &\approx &
  1.08 \times 10^7 Z/q_e^2.
  \label{sigma_mt_ep}
\end{eqnarray}
Taking into account only the dominant contribution from proton scatterings, the momentum transfer rate of electrons via the proton scatterings is given by
\begin{eqnarray}
  \Gamma_{\rm C}^{ep} &\approx& n_p \sigma_{\rm mt}^{ep}
  \label{eq_gamma_ep0} \\
  &\approx &
  \eta n_\gamma (1 -Y_\mathrm{p})
  \left[ \frac{1.08 \times 10^7}{(m_e v)^2} \right]
  \label{eq_gamma_ep1} \\
  &= & 4.42 \times 10^{9}~{\rm s}^{-1}~v^{-2}
  ~({\rm at}~T_9 =0.1).
  \label{eq_gamma_ep}
\end{eqnarray}

\subsection{Photons}
When cosmic background radiation is exposed to a plasma composed of electrons and nuclei that has a bulk velocity $v_\mathrm{out} \ll 1$, the electrons lose momenta in the direction of the bulk velocity, identified as the $z$-direction, mainly via Compton scattering off the photons. The momentum transfer rate is given as follows. First, the rate of change in the average electron momentum in the $z$-direction, $p_{e z}$, is given by
\begin{eqnarray}
  \left | \frac{dp_{e z}}{dt} \right|
  &\approx &
  \frac{1}{n_e}
  \frac{2}{(2 \pi)^3} \int d^3 p_e f_e(\bfp_e;T)
  \frac{2}{(2 \pi)^3} \int d^3 p_\gamma f(\bfp_\gamma; T) \nonumber \\
  && \times \int_{-1}^1 d \mu
  \frac{d\sigma_\mathrm{Com}}{d \mu}(\mu; \bfp_e, \bfp_\gamma)
  \frac{\int_0^{2 \pi} d \phi \Delta p_{\gamma z}(\mu, \phi; \bfp_e, \bfp_\gamma)}
       {2 \pi},
\end{eqnarray}
where
$\bfp_e$ and $\bfp_\gamma$ are momenta of an electron and a photon, respectively,
$f_e(\bfp_e;T)$ and $f(\bfp_\gamma; T)$ are the momentum distribution functions of electrons and photons, respectively,
$d\sigma_\mathrm{Com}/d\mu$ is the differential cross section of the Compton scattering as a function of $\mu =\cos \theta$ for the scattering angle $\theta$, and
$\Delta p_{\gamma z}$ is the increase of the $z$-component of the photon momentum $p_{\gamma z}$ that depends on the azimuthal angle $\phi$ at a scattering. The Pauli blocking of the electron and the induced emission of the photon have been neglected in this equation.

We assume an isotropic photon distribution in the cosmic rest frame and an isotropic electron distribution in the frame of the fluid moving in the rest frame. The electron distribution function in the cosmic rest frame is given \citep[e.g.][]{2021PhRvE.103c2101K} by
\begin{eqnarray}
  f_e(\bfp_e;T)
  &=&\frac{1}{\gamma_V}
  \left\{ \exp\left[ \frac{\gamma_V \left( E_e^\mathrm{tot} -V p_{e z} \right) }{T} \right] +1 \right\}^{-1},
\end{eqnarray}
where
$V$ is the bulk velocity of the fluid,
$\gamma_V=(1 -V^2)^{-1/2}$, and 
$E^\mathrm{tot}_e$ is the total electron energy.
When electrons are nonrelativistic and the second term in the exponent is much smaller than unity, this distribution is approximated by
\begin{eqnarray}
  f_e(p_e, \mu_e;T)
  &\approx& \frac{\exp(-\gamma_V E_e^\mathrm{tot} /T)}{\gamma_V}
  \left[ 1 +\frac{\gamma_V V p_e \mu_e}{T} \right],
\end{eqnarray}
where
$\mu_e =\cos \theta_e =p_{e z} /p_e$.
Thus, a bulk velocity $V$ induces an excess in the distribution in the direction of $V$. For example, if $V \sim 0.05$ ($\gamma_V \sim 1$) and the electron kinetic energy $E_e = E^\mathrm{tot}_e -m_e \sim T$, the excess of $\mu_e =1$ at $T_9 \sim 0.1$ is
\begin{eqnarray}
  \frac{\gamma_V V p_e}{T}
  &\sim& V \sqrt{\frac{ 3 m_e}{T}}
  \sim 0.6.
\end{eqnarray}
Then, such jets include ${\mathcal O}(10)$\% anisotropy in the distribution function of electrons.

For $T_9 \lesssim 0.1$, background photon energies are much lower than the electron mass. By neglecting the photon energy shift at the Compton scattering as a perturbation of ${\mathcal O}(E_\gamma/m_e)$, the differential cross section is reduced to that of the Thomson scattering, given by
\begin{eqnarray}
  \frac{d\sigma_\mathrm{Com}}{d \mu}(\mu)
  &\approx&
  \pi \left( \frac{e^2}{m_e} \right)^2 \left( 1 +\mu^2 \right).
\end{eqnarray}
We define the $x$-axis so that the $zx$-plane includes $\bfp_e$ The change of $p_{\gamma z}$ is then given by
\begin{eqnarray}
  \Delta p_{\gamma z}(\mu, \phi; \bfp_e, \bfp_\gamma) &=&
E_\gamma' \left[
  ( \gamma \cos \theta_e \cos \psi -\sin \theta_e \cos \phi_\gamma \sin \psi )
  \cos \theta \right. \nonumber \\
  &&-( \gamma \cos \theta_e \sin \psi +\sin \theta_e \cos \phi_\gamma \cos \psi )
  \sin \theta \cos \phi \nonumber \\
  &&\left. +\sin \theta_e \sin \phi_\gamma \sin \theta \sin \phi
  + \beta \gamma \cos \theta_e
  \right]
  -E_\gamma \gamma \cos \theta_e \cos \alpha_\gamma,
\end{eqnarray}
where
$E_\gamma$ and $E_\gamma'$ are the energies of the photon in the cosmic rest frame before and after the scattering, respectively,
$\beta$ is the electron velocity and $\gamma =(1 -\beta^2)^{-1/2}$, 
$\psi$ is the angle between $\bfp_e$ and $\bfp_\gamma^{(1)}$ that is the photon momentum in the rest frame of electron,
$\phi_\gamma$ is the angle between the $x$-axis and the $\bfp_\gamma$ projected on the $xy$-plane, and
$\alpha_\gamma$ is the angle between $\bfp_e$ and $\bfp_\gamma$.
After the integration over $\phi$, terms proportional to $\sin \phi$ and $\cos \phi$ disappear. By neglecting terms of ${\mathcal O}(E_\gamma /m_e)$ and ${\mathcal O}(\beta^2)$, the remaining quantity becomes
\begin{eqnarray}
  \Delta p_{\gamma z}(\mu; \bfp_e, \bfp_\gamma) & \approx&
E_\gamma ( 1 -\beta \cos \alpha_\gamma) \left[
  ( \cos \theta_e \cos \psi -\sin \theta_e \cos \phi_\gamma \sin \psi )
  \cos \theta + \beta \cos \theta_e
  \right] \nonumber \\
  &&
-E_\gamma \cos \theta_e \cos \alpha_\gamma.
\label{eq_dp2}
\end{eqnarray}
The angle $\psi$ satisfies:
\begin{eqnarray}
  \cos \psi &=&
  \frac{\cos \alpha_\gamma -\beta}{v_\mathrm{M}}
  \approx \cos \alpha_\gamma -\beta \sin^2 \alpha_\gamma \\
  \sin \psi &\approx &
  \sin \alpha_\gamma \left( 1 + \beta \cos \alpha_\gamma \right),
\end{eqnarray}
where $v_\mathrm{M} =\sqrt{1 -2 \beta \cos \alpha_\gamma +\beta^2 \cos^2 \alpha_\gamma}$ is the M{\o}ller velocity.
Then, after the integration over $\alpha_\gamma$, some terms including the last term in Equation (\ref{eq_dp2}) disappear, and the remaining part is given by
\begin{eqnarray}
  \Delta p_{\gamma z}(\mu; p_e, \cos \theta_e, p_\gamma) & \approx &
  \beta E_\gamma \cos \theta_e \left(
    1 - \mu
    \right).
\end{eqnarray}
The term proportional to $\mu$ is significantly canceled when it is integrated over $\mu$. Then, the first term dominates, and the anisotropic electron distribution is reflected via this term to the momentum change rate.

The rate of change in the average value of $p_{e z}$ is then estimated as
\begin{equation}
  \left| \frac{dp_{e z}}{dt} \right|
  \sim
  f_\mathrm{exc} n_\gamma(T) \sigma_\mathrm{Th}
  \left( v_{e,\mathrm{th}} E_\gamma \right),
  \label{eq_dp3}
\end{equation}
where
$\sigma_\mathrm{Th} =(8 \pi /3) (e^2/m_e)^2 =6.65 \times 10^{-25}$ cm$^2$ is the Thomson scattering cross section,
$v_{e,\mathrm{th}} =\sqrt{3T /m_e}$ is the thermal electron velocity, and
$f_\mathrm{exc}$ is the excess fraction of electron flux moving downstream given by
\begin{eqnarray}
  f_\mathrm{exc} &\sim &
  \frac{\gamma_{v_\mathrm{out}} v_\mathrm{out} p_e}{T}
  \label{eq_f_exc1} \\
  &\sim &
  \mathcal{O} (0.1) \left( \frac{v_\mathrm{out}}{0.05} \right)
  \left( \frac{T_9}{0.1} \right)^{-1/2}
  \label{eq_f_exc2}.
\end{eqnarray}

The momentum transfer rate of moving electrons via the scattering of background photons is derived with Equations (\ref{eq_dp3}) and (\ref{eq_f_exc1}) and $p_e \sim m_e v_{e,\mathrm{th}}$, as
\ba
\Gamma_\mathrm{Th}^{e \gamma}
&\sim&
\frac{|dp_{e z}/dt|}{m_e v_\mathrm{out}}
\label{eq_gamma_eg} \\
&\sim&
\frac{T}{m_e}
n_\gamma(T) \sigma_\mathrm{Th} \\
&=& 6.82 \times 10^{9}~\mathrm{s}^{-1}
\left( \frac{T_9}{0.1} \right)^{4}.
\ea
The denominator on the right-hand side of Equation (\ref{eq_gamma_eg}) is the average electron momentum in a jet with a bulk velocity $v_\mathrm{out}$.

Then, energies generated at the magnetic reconnection are quickly shared among electrons and nuclei. For example, at $T_9=0.1$ and $v_\mathrm{out} \gtrsim 0.05$, the momentum transfer rates of electrons have the ordering of $\Gamma_\mathrm{C}^{ee} \gg \Gamma_\mathrm{C}^{ep} > \Gamma_\mathrm{Th}^{e \gamma}$. Therefore, after a large number of scatterings, bulk motions are expected to realize in jets composed of electrons and nuclei associated with a minor component of dragged photons.

\subsection{dissipation scale}

The dissipation of the primordial magnetic field over a length of $L$ dissipates on a time scale of $t = 4 \pi \sigma_\mathrm{el} L^2$, where $\sigma_\mathrm{el}$ is the electrical conductivity \citep{2001PhR...348..163G}.
Then, the magnetic fields over scales smaller than the maximum dissipation scale 
$L =\sqrt{t/(4 \pi \sigma_\mathrm{el})}$ decay until a cosmic time $t$.
The conductivity is given by
\be
\sigma_\mathrm{el} = \frac{n_e e^2 \tau_\mathrm{mt}^{ep}}{m_e},
\label{eq_conductivity}
\ee
where
$\tau_\mathrm{mt}^{ep} =(\Gamma_\mathrm{C}^{ep})^{-1}$ is the momentum transfer time scale of electrons.
Using Equations (\ref{sigma_mt_ep0}) and (\ref{eq_gamma_ep1}), the maximum diffusion scale is transformed to
\ba
L &=& \left( \frac{m_e t \Gamma_\mathrm{C}^{ep}}{4 \pi e^2 n_e } \right)^{1/2} \\
&=& \left( \left( \frac{m_e}{4 \pi e^2} \right)
  \left\{ \frac{2^2 \pi \alpha^2}{v^2 q_e^2}
  \left[ 1-\mu_{\rm max}(T) \right]^{-1/2} \right\}
  \left( \frac{3\sqrt{5} m_{\rm Pl}}{ 4 \pi^{3/2} g_\ast^{1/2} T^2} \right)
  \right)^{1/2} \\
&\sim & \left[ \frac{\sqrt{10}}{8 \pi^{3/2} }
  \frac{m_e  m_{\rm Pl} \lambda_\mathrm{D}(T)}{ g_\ast^{1/2} T^3}
  \right]^{1/2} \\
&=& \left\{ \frac{\sqrt{5}}{16 \pi \left[ \zeta(3) \alpha \right]^{1/2}}
    \frac{m_e  m_{\rm Pl} }{ g_\ast^{1/2} T^4
    \left[ \eta \left( 2 -Y_\mathrm{p}/2 \right)
      \right]^{1/2} }
    \right\}^{1/2},
\ea
where
$m_\mathrm{Pl}$ is the Planck mass, and
$g_\ast =2 +(21/4) (4/11)^{4/3}$ is the effective statistical degrees of freedom after the $e^+$--$e^-$ annihilation.
The Debye length $\lambda_\mathrm{D}$ increases with decreasing $T$, and the maximum diffusion scale monotonically increases with decreasing $T$.
At $T_9 =0.1$, this scale corresponds to
$L = 18.4~\mathrm{km}~( T_9 / 0.1 )^{-2}.$

\subsection{Reconnection rate}
The reconnection rate in the Sweet-Parker model \citep{1958IAUS....6..123S,1957JGR....62..509P} is given by
\ba
\Gamma_\mathrm{rec} &=&
\frac{v_\mathrm{in}}{L} 
=\frac{V_\mathrm{A} /L}{\sqrt{S}} 
=\left( \frac{V_\mathrm{A} \eta_\mathrm{mag}}{L^3} \right)^{1/2},
\ea
where
$v_\mathrm{in}$ is the inflow velocity of magnetic field lines into the reconnection region,
$L$ is the size of the reconnection region, and
$S$ is the Lundquist number defined as
$S= V_\mathrm{A} L/\eta_\mathrm{mag}$,
where
$V_\mathrm{A}$ is the Alfv\'{e}n velocity, and
$\eta_\mathrm{mag} =1 /(4 \pi \sigma_\mathrm{el})$ is the magnetic diffusivity.

In the epoch after the $e^+ e^-$ annihilation, electrons are dominantly scattered via Coulomb scattering by nuclei, except for self-scatterings. For example, at $T_9=0.1$, the thermal electron velocity is $v =\mathcal{O}(0.1)$, and the momentum transfer rate is $\Gamma_\mathrm{C}^{ep} \sim 10^{11}$ s$^{-1}$ (Equation (\ref{eq_gamma_ep})). Under this condition, by using Equations (\ref{sigma_mt_ep}), (\ref{eq_gamma_ep0}), and (\ref{eq_conductivity}), the reconnection rate is given by
\ba
\Gamma_\mathrm{rec}
&=& \left( \frac{V_\mathrm{A} }{L^3} \frac{1}{4 \pi}
\frac{m_e \Gamma_\mathrm{C}^{ep}} {n_e e^2}
\right)^{1/2} \\
&\sim & \left( \frac{V_\mathrm{A} }{4 \pi \alpha} \frac{m_e \sigma_\mathrm{mt}^{ep}}{L^3}
\right)^{1/2} \\
&\sim & \left( \frac{V_\mathrm{A} }{4 \pi \alpha} \frac{1.08 \times 10^7}{m_e v^2 L^3}
\right)^{1/2}.
\label{gamma_rec2}
\ea
The Alfv\'{e}n velocity for a relativistic plasma is given by
\ba
V_\mathrm{A} &=& \frac{B}{\sqrt{4 \pi \left(
    \rho_\mathrm{EM} +P_\mathrm{EM} \right)}} \\
&\approx &
\sqrt{\frac{3 f_B}{2 f_\mathrm{EM}}},
\label{alf1}
\ea
where
$B$ is the magnetic field amplitude,
$\rho_\mathrm{EM}$ and $P_\mathrm{EM}$ are the total energy density and pressure of electromagnetically coupled plasma in jets, respectively,
  $f_B =\rho_B/\rho_\mathrm{rad}$ is the ratio of the magnetic field energy density $\rho_B =B^2 /(8 \pi)$ to the total radiation energy density $\rho_\mathrm{rad}$, and
$f_\mathrm{EM} =\rho_\mathrm{EM} /\rho_\mathrm{rad}$.

  The dissipated energy is shared dominantly by electrons and nuclei. Then, the total energy in plasma jets is dominated by moving baryon energy. The energy density of the plasma is approximated by
$\rho_\mathrm{EM} \simeq m_\mathrm{b} \gamma_A n_\mathrm{b}$,
where
$m_\mathrm{b}$ is the baryon mass, and
$\gamma_A =(1 -V_A^2)^{-1/2}$.
Then, it follows that
\ba
f_\mathrm{EM} &=&
\frac{60 \zeta(3)}{ \pi^4 g_\ast} \frac{m_\mathrm{b}}{T} \eta \gamma_A
\label{eq_f_EM0}
\\
&=& 1.44 \times 10^{-5}~\left( \frac{T_9}{0.1} \right)^{-1}
\left( \frac{\eta}{6 \times 10^{-10}} \right) \gamma_A.
\label{eq_f_EM}
\ea
Therefore, if the field amplitude is as high as $f_B \sim 10^{-4}$--$10^{-3}$, the jets are highly relativistic with $V_A \sim 1$ and $\gamma_A \gg 1$.

Using Equation (\ref{gamma_rec2}) and $V_A \sim 1$, the reconnection rate is derived as
\ba
\Gamma_\mathrm{rec}
&=& 2.03~{\rm s}^{-1}
v^{-1}
\left( \frac{L}{10~{\rm km}} \right)^{-3/2}.
\label{eq_rec_rate}
\ea

During the reconnection, jets involve surrounding background plasma, and bulk motions over large volumes would form. We roughly estimate a relation of the bulk velocity $v_\mathrm{out}$ and the reconnection region range $L$ as follows. By the reconnection, the magnetic field energy in the volume of $\sim L^3$ is converted to the energy of plasma jets ranging over a scale of $l_\mathrm{jet} \sim v_\mathrm{out} /\Gamma_\mathrm{rec}$. As a result, there is an energy conservation for nonrelativistic jets produced by the reconnection:
\ba
\rho_B L^3
=\frac{\rho_\mathrm{EM} v_\mathrm{out}^2}{2} l_\mathrm{jet}^3
=\frac{\rho_\mathrm{EM} v_\mathrm{out}^5}{2 \Gamma_\mathrm{rec}^3}.
\label{eq_conservation}
\ea
Then, it follows with Equations (\ref{eq_f_EM}) and (\ref{eq_rec_rate}) that
\ba
v_\mathrm{out} &\sim &
\left[ \frac{2 \rho_B}{\rho_\mathrm{EM}} \left( \Gamma_\mathrm{rec} L \right)^3 \right]^{1/5} \\
&=& \left\{ \frac{2 f_B}{f_\mathrm{EM}} \left[ 20.3~\mathrm{km}~\mathrm{s}^{-1} v^{-1} \left( \frac{L}{10~{\rm km}} \right)^{-1/2} \right]^3 \right\}^{1/5} \\
&\sim & 8.45 \times 10^{-3} 
\left( \frac{f_B}{10^{-3}} \right)^{1/5}
\left( \frac{T_9}{0.1} \right)^{1/5}
\left( \frac{\eta}{6 \times 10^{-10}} \right)^{-1/5}
\left( \frac{L}{10~{\rm km}} \right)^{-3/10}.
\label{eq_final_v}
\ea
Note that it is assumed that baryons in such extended jets are nonrelativistic. Thus, when the reconnection completes, the bulk velocity is small. During the reconnection, small and relativistic jets form first, and they gradually become large and slow. Nonthermal nucleosynthesis operates until the kinetic energies of nuclei in the jets become insufficient for inducing reactions.

\subsection{Simple model}
\begin{figure}[h]
  \centering
  \includegraphics[width=5in]{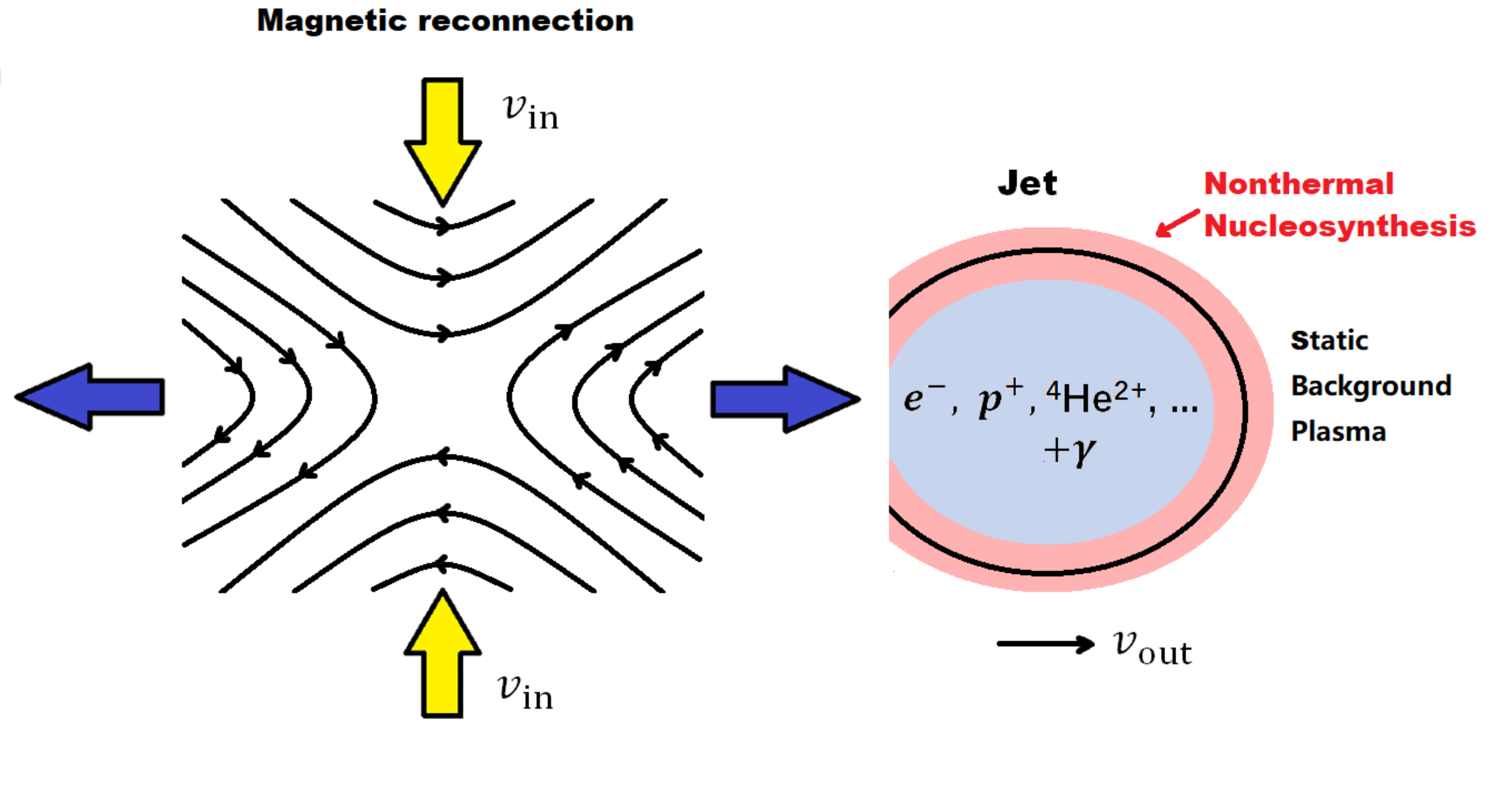}\\
  \caption{Model of the cosmic ray nucleosynthesis triggered by the magnetic field reconnection. Magnetic reconnection converts the magnetic energy to the kinetic energy of charged particles. Because of the coupling of electrons, nuclei, and photons, plasma jets with bulk velocity $v_\mathrm{out}$ are formed. Nonthermal nucleosynthesis takes place where the jets are thermalized at their boundaries.
  }
  \label{fig_app1}
\end{figure}

Figure \ref{fig_app1} shows a schematic picture of the cosmic ray nucleosynthesis triggered by large-scale magnetic field reconnection, inferred from the estimates given above. Tightly coupled plasma of electrically-charged electrons and nuclei partially drags background photons at $T_9 \sim 0.1$. Therefore, jets composed of electron-nuclei-photon fluid with a bulk velocity form. Depending on the magnetic field energy density, the kinetic energies of nuclei in the jets can be large enough that nonthermal nucleosynthesis effectively occurs when the jets collide with static background plasma.

\bibliography{ref1}{}
\bibliographystyle{aasjournal}





\end{document}